# The Non-linear Dynamics of Meaning-Processing in Social Systems

Loet Leydesdorff

Amsterdam School of Communications Research (ASCoR), University of Amsterdam, Kloveniersburgwal 48, 1012 CX  Amsterdam, The Netherlands.

<loet@leydesdorff.net> ; <http://www.leydesdorff.net>

**Abstract**

Social order cannot be considered as a stable phenomenon because it contains an order of reproduced expectations. When the expectations operate upon one another, they generate a non-linear dynamics that processes meaning. Specific meaning can be stabilized, for example, in social institutions, but all meaning arises from a horizon of possible meanings. Using Luhmann's (1984) social systems theory and Rosen's (1985) theory of anticipatory systems, I submit equations for modeling the processing of meaning in inter-human communication. First, a self-referential system can use a model of itself for the anticipation. Under the condition of functional differentiation, the social system can be expected to entertain a set of models; each model can also contain a model of the other models. Two anticipatory mechanisms are then possible: one transversal between the models, and a longitudinal one providing the modeled systems with meaning from the perspective of hindsight. A system containing two anticipatory mechanisms can become hyper-incursive. Without making decisions, however, a hyper-incursive system would be overloaded with uncertainty. Under this pressure, informed decisions tend to replace the "natural preferences" of agents and an order of cultural expectations can increasingly be shaped.

# 1. Introduction

"Society does not consist of individuals, but expresses the sum of the relations within which individuals stand" (Marx [1858], 1973, at p. 265). Marx was the first to note that this system of social relations transforms both nature and history. From his perspective this transformation is considered as "alienation" under the condition of capitalism. Bhaskar (1979; 1998, at p. 207) emphasized that a transformational model of social activity entails that the subject-matter of the social is composed of relations. The relations can be considered as links in networks of relations. The networks develop over time with the relational links as the operators. By evolving, however, these networks also transform themselves, their relevant environments and pasts (Maturana, 1978).

Husserl (1929) theorized that the transformative power in networks of interhuman relations can be considered as *intersubjective* intentionality. The substance of this "intersubjectivity" is different from subjective intentionality; Husserl (*ibid.*, p. 144) calls it, with reference to Leibniz, another "monad." One has no access to this domain other than by a phenomenological reduction that places the subjective intentions between brackets so that the intersubjective domain transcends the individual one. According to Husserl, the study of this domain would provide us with "a concrete ontology and a theory of science" (*ibid.*, at p. 159). However, the author conceded that he had no instruments beyond this transcendental apperception of the domain and therefore had to refrain from empirical investigation:

> We must forgo a more precise investigation of the layer of meaning which provides the human world and culture, as such, with a specific meaning and therewith provides this world with specifically "mental" predicates. (Husserl, 1929, at p. 138; my translation).

Following Husserl's notion of intentional "intersubjectivity," Luhmann (1984) elaborated on the concept that networked social systems are not only transformed by human actions, but additionally—and according to Luhmann's definitions even exclusively—able to transform themselves *reflexively*. In his opinion, "intersubjectivity" had introduced into sociological theorizing a semantic emphasis on what human beings share (Schutz, 1932, 1952; Habermas, 1981; cf. Grant, 2003; Leydesdorff, 2007). Luhmann considered human actors ("consciousness systems") only as the environment of the social system. Therefore, he proposed replacing Husserl's concept of "intersubjectivity" with communication-theoretical concepts like "interaction," "organization," and "self-organization" (Luhmann, 1994). Communication-theoretical terms provide a different starting point because of the emphasis on potential surplus in the non-linear interaction terms (Grathoff, 1978).

In the "double contingency" of *interactions*, for example, *Ego* knows the *Alter* to be a reflexive *Cogito* who equally entertains expectations (Parsons, 1968; Elmer, 1995; Vanderstraeten, 2002). Thus, a social reality is constructed in which a symbolic order is invoked (Lévi-Strauss, 1987). Interactions can be face-to-face or organized into systems of communication. Because reflexive communications are the operator of the social system, one can expect that the *organization* of systems of communication will endogenously be transformed over time. This *self-organization* of the communication can be expected to add a third coordination mechanism to networked interactions and the organization of communications (Luhmann, 1975a, 2000).

Note that the analogy with subjective reflexivity remains formal because the operation is substantively different: while individuals process thoughts, perceptions, and feelings, social systems operate only in terms of communications. In both cases, meaning is generated and communicated, but one can expect the dynamics and control mechanisms to be different. The two systems (the social and the individual) can also be expected to remain reflexive on developments in each other. Luhmann (1986) used Parsons's (1968) concept of "interpenetration" and Maturana's (1978) "structural coupling" in order to understand this mutual plasticity in the construction and processing of meaning (Luhmann, 1978a).[1]

---

[1] Structural coupling between consciousness and communication as systems relate them as each other's relevant environments. "Intepenetration" adds an operational coupling between these two systems since meaning-processing systems have reflexive access to each other's operations (Luhmann, 1988, 1991).

For example, by codifying communication into scientific discourse a paradigm can sometimes be developed at the supra-individual level. Individuals participating in the communication are reflexively able to update the meaning of their previously held concepts and they can also generate knowledge-claims—that is, variation—to be selected or deselected by the codified discourse. However, a scientific paradigm can only be changed as a result of selections that function at the supra-individual level of scientific discourse (Kuhn, 1962; Mulkay *et al*., 1983; Gilbert & Mulkay, 1984). Following Parsons (1951), Luhmann (1984) added that codified meanings can be "condensed" in subsystems that are *functionally* differentiated. The functionality of the differentiation means that the system is able to use the specificity of the code for its reproduction. Using the various codes, events can then be appreciated differently within each subsystem; this diversity enables the system to process more complexity and, therefore, can be considered as functional to the further development of the system. The communications can increasingly be refined.

Künzler (1987, at p. 323) noted that Parsons considered codes as specializations of language, while Luhmann (1975b) emphasized an additional role of codes of communication under the condition of functional differentiation: the codes span functional subsystems as structures of communication. These different structures provide selection mechanisms to one another. Different meanings can be further selected by different codes of communication. For example, knowledge can be considered as a meaning that makes a difference. Scientific knowledge is generated and reproduced in highly codified discourses, but at the level of the social system (Mulkay *et al*., 1983; Leydesdorff, 2006a).

Parsons (1963a and b) introduced "symbolic generalization" of the code of communication to explain the sometimes binding character of collective action. His concept remained close to Weber's ([1904], $^3$1968) concept of values, except that the values are no longer considered as given; symbolic generalization functions as a mechanism for integrating the social system despite its distributed—that is, subsymbolic—mode of operation (Leydesdorff, 1993). Using money as the model of an exchange medium, Parsons generalized this concept of media to include power and influence as other exchange media that can be generalized symbolically.

Luhmann (1975) elaborated on these concepts by considering power, etc., as codes of communication, each of which can be generalized in other directions under the condition of functional differentiation. In a previous hierarchical order, the different media were subsumed under each other so that a cosmological order could eventually be fixed (Toulmin, 1990; Leydesdorff, 1994). For example, the Emperor had to go to Canossa because the Pope had eventually the power to "ex-communicate" him. Under the condition of modernity, however, exchange media can develop according to their own logic, and thus science, politics, the economy, affection, etc., can develop their specific codes of communication further along (analytically!) orthogonal axes. Symbolic generalization implies that every event can be assessed from the specific perspective of each code of communication. Luhmann (1984, 1997) assumed that these functional systems are "operationally closed" or, in other words, that the codifications are domain-specific and not possibly interacting.

For example, everything can be assessed in terms of its economic value or its esthetic beauty, and these assessments can no longer be expected to correspond because they are coded differently. Although Luhmann paid little attention to the further development of the formalization, he considered formalization as a necessary next step (Baecker, 1999, 2005). For example, Luhmann (1990, at p. 340) argued that differentiation in the codification generates a feedback that changes the social system, but he added that "developing this perspective is only possible if an accordingly complex systems-theoretical arrangement is specified."

How can such "an accordingly complex systems-theoretical arrangement" be constructed? How can one move from Luhmann's theoretical reflections to a model specified in terms of systems operations? Would the specification of a model also enable us to run simulations? (Kron, 2002). Luhmann (1984, at p. 110, [1995, p. 73]) declined Habermas's (1981, at pp. 152 ff.) attempts to ground "meaning" *methodologically* at the level of individuals by using action theory. The "ccordingly complex systems-theoretical arrangement" should enable us, among other things, to reconstruct time within the social system (Luhmann 1984, at pp. 116 ff. [1995, at pp. 78 ff.]). From Luhmann's perspective, this grounding of reflexivity in the distinction among different time axes would raise interesting issues about the epistemological relations between social systems theory, biological theorizing, and general systems theory (Luhmann, 1990; Cornell, 1992).

In this study, I submit that the mathematical theory of anticipatory systems (Rosen, 1985) provides us with a model of anticipation that can be elaborated into a theory of intentionality at the supra-individual level, and into methods for the simulation using Dubois's (1998) computational algorithms of *incursivity*. Rosen's theory of anticipatory systems and Dubois's new algorithms allow us to consider time as a degree of freedom within systems. Meaning is provided from the perspective of hindsight. In a previous study (Leydesdorff, 2005), I showed how an observer can be generated *within* a system using the incursive formulation of the logistic equation. The observer entertains a model of the modeled system within the system. Leydesdorff & Dubois (2004) proved that anticipation in interactions between and aggregations of social agents can be modeled using this same algorithm.

In these (biologically-inspired) models, neither the agents nor the resulting social systems were endowed with the capacity to use these models reflexively. In this study, I shall show that under the condition of functional differentiation among the codes of communication a social system can be expected to contain two analytically independent axes for the anticipation: a transversal and a longitudinal one. The interaction among these two mechanisms of anticipation can be used to generate a set of three *hyper-incursive* equations. These equations can be provided with sociologically meaningful interpretations. One of these equations will enable us to understand why a social system can be considered as strongly anticipatory. Weakly anticipatory systems are only able to entertain models of their future, but not to reconstruct the modeled system from the perspective of the model, that is, by using the reverse arrow of time for an additional reflection.

## 2. Anticipation, meaning, and intentionality

Rosen (1985) defined anticipatory systems as systems that contain models of themselves. The modeling part advances on the modeled one and provides meaning to the system from the perspective of hindsight. Thus, an anticipatory system generates internally a time difference ($\Delta t$) between the model and the modeled system. Rosen himself provided examples of how biological systems can sometimes anticipate, but without intentionality, that is, without room for choice and under Darwinian selection pressure. Meaning-*processing* systems, however, are able to entertain more than a single model of themselves. If a multitude of models can be maintained, the set of models can reflexively be considered as a horizon of possible meanings, from which the system has to make an additional selection before a historical instantiation can be constituted (Giddens, 1981 and 1984).

The reflexive selection from a horizon of meanings can also be considered as intentionality. In other words, an intentional system has an additional degree of freedom—the variety of meanings—from which it selects reflexively before it can organize itself in a historical manifestation. When meanings can be processed within a system using an internal degree of freedom for the reflection, other possible meanings can also be appreciated. In other words, intentional openness emerges. In the case of biological systems, meaning can also be generated (Maturana, 1978), but it is selected deterministically by the environment in a process of natural selection or over time because of the progression of a naturally determined life-cycle.

Let us now further distinguish between social systems and individuals: decision-making at the individual level can be expected to lead to an *integration* of expectations, while in the communicative mode of the social system the process may remain *differentiated* both at each moment and over time. Luhmann (e.g., 1984, at p. 625 [1995, at p. 461]; 1994) proposed at some places considering the social system of communications as a "dividuum:" a variety in the forms of integration in meaning processing can be expected.

For example, policies from an economic perspective may be different from decisions taken from a scientific perspective. However, we shall see that decision-making remains crucial because it enables us (or, more generally, meaning-processing systems) to reduce the uncertainty when interfacing differently coded expectations.

Meaning-processing systems develop both historically—that is, with the arrow of time—and against the arrow of time, by providing meaning to their past from the perspective of the present and with reference to future options of further development. The future options can be considered as a horizon of meanings from which the current meaning is selected. However, meaning is also selectively provided to historical events. In other words, various subdynamics of meaning can be distinguished in terms of their external references. When meaning is provided to a historical given, this is done with hindsight; but if meaning is provided to another meaning, the historical order is no longer given: different perspectives may use different time horizons.[2] In the case of an innovation, for example, one may wish to combine short-term market perspectives with longer-term technological ones.

Some models may advance more rapidly than others. For example, the market can be expected to operate faster than a research process, and interface-management in a knowledge-based corporation is therefore a difficult task in terms of assessments of time-horizons: "which technology can be introduced on the market at which moment in the future?" The models are synchronized at $t = t$ by decisions at interfaces. This synchronization is based on an expected time difference at the interface. The time differences between and among systems will be different from the time differences in the updates *within* each of the differently codified processes of communication. In other words, the subsystems use different clocks (Leydesdorff, 1994). Accordingly, the expectations vary with the system of reference, and priority among them has to be constructed in historical time. The reduction of uncertainty by a decision changes in turn

---

[2] In German, the difference between two senses of meaning can be expressed using the words "Bedeutung" and "Sinn." Luhmann (e.g., 1978b, at p. 56) defined "Sinn" as follows: "Etwas hat Sinn, wenn es Verweisungsüberschüsse fixiert" (Something has meaning when it fixes a surplus of references). Using this definition, the two meanings distinguished above are kept together under a single and foundational category (Schützeichel, 2003, at p. 35, note 1.).

the relevant horizon of meanings, and thus decisions keep the cycles going (Luhmann, 2000).

These transversal selections of meaning among subsystems are additional to the longitudinal development of meaning (e.g., updates) within self-referential (sub)systems. The longitudinal selection is based on Rosen's model, but the transversal asynchronicity is added on the basis of Luhmann's model of functional differentiation in the codification of meaning. Note that the longitudinal selection is sequential and hierarchical, while the transversal one can be expected to change with the nature of the differentiation. The two axes stand orthogonally under the condition of functional differentiation. In other words, the longitudinal generation of meaning was already available as an incursive mechanism in pre-modern societies, while transversal updates provide modern societies with an analytically independent mechanism for organizing reflexive communications in an anticipatory mode.

Let me recapitulate the *three* different selection mechanisms involved in meaning-processing:
1. First, there is a selection mechanism between a system and its environment as in biological systems. However, the reflexive system is able to generate models of its environment by selecting upon the uncertainty in the events, while a biological system is only selected passively for survival by the environment. Note that selections are asymmetrical operations;
2. Second, when a variety of meanings can be entertained, some first-order selections are selected for stabilization ("condensation"); meaning is anchored in meaning-processing systems by codification;
3. Third, when differently codified meanings are interfaced, some stabilizations can also be selected for globalization. In other words, a complex dynamics of meaning-processing is generated.

Selections from global horizons of meaning add this third selection mechanism to momentary selections from historical events in the present, and ongoing stabilizations of meaning within meaning-processing systems. However, species of chaotic behaviour can be generated when three subdynamics operate upon one another (May & Leonard, 1975). In other words, the social system can be expected to evolve as a non-trivial machinery *because* it contains a complex—that is, composed and non-linear—dynamics of meaning-processing (Baecker, 2002). The alignment of expectations in social order can no longer be taken for granted, but would need always to be explained (Hobbes, 1651; Luhmann, 1995a).

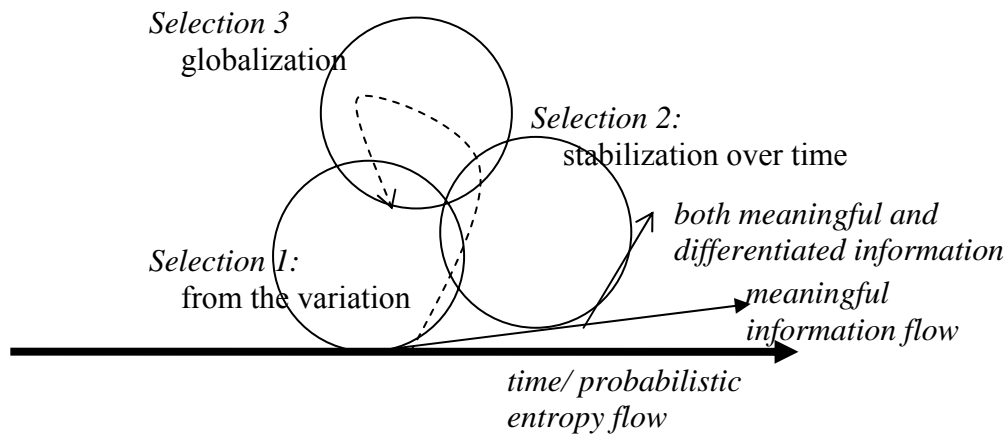

**Figure 1**: Three selections cycle in the case of processing meaning on top of the (Shannon-type) information flow.

I have tried to visualize the non-linear configuration in Figure 1: First, the historical variation is reorganized by the selecting system into a flow of meaningful information. By ongoing codification processes some meanings are discarded and others retained in a second round of selections of meaning. If the system has one more degree of freedom for further selections, a "triple helix" of meaning-processing can be generated (Lewontin, 2000; Leydesdorff, 2006b). This non-linear feedback on the underlying information flow may lead to the latter's reorganization.

One would expect the results of the three subdynamics to be different, since selections operate asymmetrically. The dashed arrow is intended to represent the resulting behaviour which may lead to feeds forward at some places, but backward at others. At the level of the individual identity, the feeds forward and feedbacks have continuously to be integrated in "real life." However, after its functional differentiation, the social system can no longer be conceptualized as an "individuum." While a stratified system like a high culture can be expected to contain a center of control, a functionally differentiated one may also fail to integrate: the different integrations of meaning by individuals and organizations remain distributed. The uncertainty contained in this distribution can be used as second-order variation along another (orthogonal) axis. Second-order selection along this axis would feed back on previous stabilizations as "globalization" (cf. Hayami & Rutan, 1970; Nelson & Winter, 1982, at p. 57).

In Luhmann's social systems theory, the globalizing horizons of meaning tend to remain unspecified otherwise than with reference to Husserl's transcendental phenomenology (e.g., Luhmann, 1995b). However, the functional differentiation among the subsystems may also lead to *different* sets of anticipations. Luhmann (1984, at p. 605 [1995, at pp. 446f]) mentioned "recursive anticipation" as a condition for the autopoietic reproduction of the complex arrangement. The anticipatory mechanisms, however, were not specified in terms of their differentiation along the time axis. Meaning which is provided to past events can be expected to exhibit a dynamic that differs from meaning assigned to future possibilities.

**3. Incursion**

Two sets of equations help to make my argument. The first is taken from Andersen (2002, at pp. 170 ff.) who defined the logistic equation—the second set of equations to be discussed below—as a special case of the *perturbed recursion model.* Let me introduce this model using a first set of equations (Figure 2):

$$S_t = F(S_{t-1}, P) \tag{1}$$

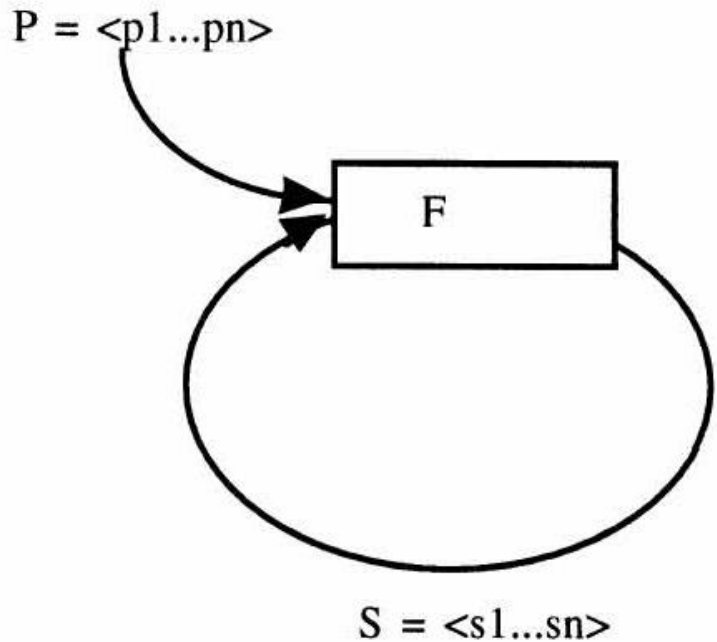

**Figure 2**: the perturbed recursion model; $S_t = F(S_{t-1}, P)$ (Source: Andersen, 2002, at p. 170).

$F$ is a recursive function that transforms state $S_{t-1}$ into a new state $S_t$, using a set of parameters $P = p_1, ..., p_n$. Baecker (2002, at pp. 86 ff.) argued that in the case of Luhmann's model of self-organization, the function ($F$) can be considered an operator of the communicatin system ($S$) itself, and used the word "eigen-function" for this operation. However, different eigenfunctions can be expected to structure variation differently. In the case of communication networks, the recursive selections can be expected to codify different meanings increasingly along the main axes of the network.

One can add a subscript $c$ (of *coding*) to the operation and use Baecker's (2002) formalization of the system/environment distinction, as follows:

$$S_t = F_c(S_{t-1}, E) \qquad (2)$$

When functionality of the differentiation prevails, the external environment $E$ is composed for each subsystem of other subsystems, with a remaining term $\varepsilon$ as representation of the residual environment. I shall use the lower-case $f$ and $s$ for this level, and then rewrite Equation 2 as follows:

$$s_{i,t} = f_i(s_{i,t-1}, s_{j,t-1}, s_{k,t-1}, s_{l,t-1},..., \varepsilon) \qquad (3)$$

In a functionally differentiated system, the windowing of the subsystems upon each other becomes horizontal. Each subsystem ($i$) codes ($f_i$) its own previous development and the development which it finds in its relevant environments. While the relation to the subsystem itself proceeds historically, that is, from its state at $t-1$ to its state at $t$, meaning can be provided to this development by the other (selecting) subsystems from the perspective of hindsight. For example, the development of one subsystem of society (e.g., a technology $s_{i,t}$) contains both a reference to a previous state of this technology ($s_{i,t-1}$) and a reference to a current state of the selecting subsystem, e.g., the market ($s_{j,t}$) (Sahal, 1981). Thus, we obtain the additional equation:

$$s_{i,t} = f_i(s_{i,t-1}, s_{j,t}, s_{k,t}, s_{l,t},..., \varepsilon) \qquad (4)$$

The state of a subsystem ($s_{i,t}$)—and, therefore, of the system—is now dependent both on the previous state of this subsystem ($s_{i,t-1}$), and the current states of other subsystems of the system. Dubois (1998) proposed labeling an evaluation that includes the current state "incursion" in order to distinguish it from "recursion" to a previous state. In a later section, I shall use the concept "hyper-incursion" for an evaluation that includes future states.[3]

---

[3] Dubois (1998b, at p. 208f.) defined "hyperincursion" as an incursion that generates multiple future states.

Before extending this equation in the next section, let me first elaborate on the notion of "incursion" using the logistic equation. The logistic equation is relevant because a self-referential system considers an environment (*E*) as *its* carrying capacity. The environment can therefore be defined only with reference to the system itself as its complement (Spencer Brown, 1969). The logistic equation normalizes this definition of the environment as a relation between a system (*x*) and its complement (1 – *x*) as follows (Devaney, 2003):

$$x_t = ax_{t-1}(1 - x_{t-1}) \quad ; \quad 0 \leq x \leq 1 \tag{5}$$

The selection environment $(1 - x_{t-1})$ inhibits further growth of the system as a feedback term. For *x* = 1, the system is selected away. For relatively small values of the parameter *a* (1 < *a* < 3), this model generates the well-known sigmoid curves of biological systems which grow and go through transitions because of increasing selection pressure. (For larger values of *a*, the model bifurcates into an oscillation (at $a \geq 3$) or increasingly generates chaos (3.57 < *a* < 4).)

Unlike this biological (e.g., population-dynamical) model which operates as selection pressure on a system and transforms during its development along the arrow of time under natural selection pressure, meaning is provided by the system from the perspective of hindsight. This can be modeled using the logistic equation, but with different time parameters as follows (Dubois, 1998):

$$x_t = ax_{t-1}(1 - x_t) \tag{6}$$

In this *incursive* case, the system builds as variation on its previous state, but the feedback term provides the selection environment in the present. Without any addition, Equation 6 can now be reorganized analytically into Equation 7 as follows:

$$x_t = ax_{t-1}(1 - x_t) \tag{6}$$

$$x_t = ax_{t-1} - ax_{t-1}x_t$$

$$x_t + ax_{t-1}x_t = ax_{t-1}$$

$$x_t(1 + ax_{t-1}) = ax_{t-1}$$

$$x_t = ax_{t-1}/(1 + ax_{t-1}) \tag{7}$$

Unlike the recursive formulation of the logistic equation, this system cannot generate chaotic phenomena because it contains a steady state to which it converges, for all values of the parameter ($a$).[4] As noted, this model allowed us to generate an observer—that is, a modeling subsystem—in the simulation (Leydesdorff, 2005). The challenge, however, remains the construction of an observing system that is able to predict—or, in other words, to reduce uncertainty about—developments in the complex system under observation (Dubois, 2002).

**4. Functional differentiation**

Let us return to Equation 3 and 4 above for the derivation of how to combine functional differentiation with incursive algorithms.

$$s_{i,t} = f_i(s_{i,t-1}, s_{j,t-1}, s_{k,t-1}, s_{l,t-1}, ..., \varepsilon) \tag{3}$$

$$s_{i,t} = f_i(s_{i,t-1}, s_{j,t}, s_{k,t}, s_{l,t}, ..., \varepsilon) \tag{4}$$

Note that all systems and subsystems continue to operate historically and thus provide meaning to their own development along the longitudinal axis at the same time as they provide meaning to one another. There is no prescribed order: relations between

---

[4] This system has a steady state for all values of $a$. The steady state can be found by solving $x_t = x_{t-1}$ as follows:

$$x = ax/(1+x) \tag{7a}$$
$$x(1+ax) = ax$$
$$ax^2 + (1-a)x = 0$$
$$x = 0 \ \lor \ x = (a-1)/a$$

subsystems can be expected to develop both incursively and recursively. In other words, both Equation 3 and Equation 4 are continuously operational. However, they differ in some respects with a Δ*t*. In other words, the functionally differentiated system contains an additional Δ*t* which can be used for a reversal of the time axis at local interfaces and thus generate a transversal incursion which stands orthogonally to the longitudinal incursions that were discussed in the previous section.

The transversal incursion can be formalized in a manner analogous to that of the incursive formulation of the logistic equation provided in Equation 5, but with additional references to different subsystems:

$$x_{i,t} = ax_{i,t-1}(1 - x_{j,t}) \tag{8}$$

On the basis of this model one can generalize Equation 9 for cases where other subsystems provide also relevant selection environments for the development of subsystem *i*, as follows:

$$x_{i,t} = ax_{i,t-1}(1 - x_{j,t})(1 - x_{k,t})(1 - x_{l,t})\ldots\varepsilon \tag{9}$$

Each subsystem (*i*) develops with reference to its own previous state, but one can expect all other subsystems to feedback upon this development by entertaining a model of the reference system in the present using their own respective codes. Since each meaning-providing subsystem (*i*) also provides meaning to its own development longitudinally, and the two types of meaning-providing can interact, one can rewrite Equation 9 as follows:

$$x_{i,t} = ax_{i,t-1}\prod_{j=1}^{n}(1 - x_{j,t}) \cdot \varepsilon \tag{10}$$

In this formula *n* represents the number of subsystems of the functionally differentiated system. While this number was analytically restricted in Parsons's structural-functionalism—using his so-called four-function paradigm—the number of subsystems can vary in Luhmann's (1997) theory with the historical development of the media of communication and their symbolic generalization into codes. When the various subsystems use different frequencies for the updates, parameters have to be added to the corresponding selection mechanisms. The importance of these potentially different speeds—and the accordingly different parameters—for the further development of the system will become manifest in the next section.

**5. Stabilization, Meta-stabilization, and Globalization**

Let us now turn to the question of what one can expect when a given code operates upon a differently codified meaning at one or more interfaces. I mentioned above that the operation of two differently codified incursions upon each other may lead to "mutual shaping" and the consequent stabilization of a co-evolution along a trajectory. The formalization will enable us to distinguish between stabilization in a co-evolution between two subdynamics and the possibility of globalization in the case of three subdynamics. On the basis of this distinction, we will be able to appreciate the hyper-incursive formulas of the next section in communication-theoretical terms.

When selection is represented by the feedback term of the logistic equation, that is, by $(1 - x)$, two different selections, operating as subdynamics of the same historical variation $(ax)$, would result in a *selection pressure* prevailing at the systems level of $(1 - bx)(1 - cx) = bcx^2 - (b+c)x + 1$. (The specificity of the selections is expressed as parameters *b* and *c*, representing relative selection weights.) For positive values of *b* and *c* (that is, both selections are operating),[5] the resulting selection environment would no longer be linear, but curved with a minimum. In other words, the selection environment of a system containing two independent selection mechanisms can be represented as a quadratic curve with a minimum (Figure 3a).

---

[5] For $b = 0$ or $c = 0$, the selection is no longer operative; the model degenerates to the previous (simple) case. If one of the parameters changes in sign, the system becomes meta-stable, and a bifurcation can be expected. For example, the change in sign can be appreciated as a market with decreasing marginal return versus one with increasing marginal returns. The latter market is meta-stable and can be expected to lead to bifurcation and lock-in (Arthur, 1988, 1989; Leydesdorff & Meyer, 2006).

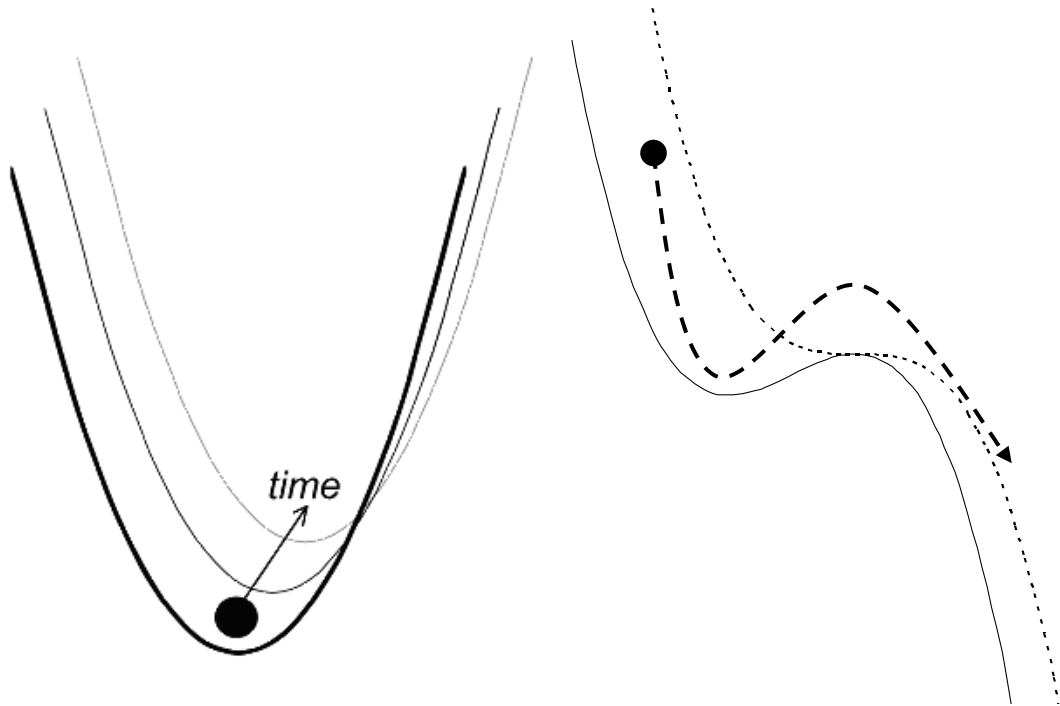

**Figure 3a**:
——— $f(x) = v (x^2 - 2x + 1)$ (stabilized)
($b = 1$ and $c = 1$)

**Figure 3b**:
------ $f(x) = -v (x^3 - 3x^2 + 3x - 1)$
——— $f(x) = -v (hx^3 - ix^2 + jx - k)$
(Eq. 11)

When this minimum is extended along the time dimension, a valley is shaped in which the system can follow a trajectory (Sahal, 1985; Waddington, 1957). Since an evolving system can be expected to generate variation, the system does not have to be stabilized at the minimum, but may, for example, oscillate around it. Note that I model here only the shape of the selection environment; multiplication with the variation term ($v = ax$) normalizes the system within the unit interval.

Adding one more selection term to the selection environment leads analogously to the following equation:

$$f(x) = (1 - bx)(1 - cx)(1 - dx)$$

or equivalently (but using different parameters):

$$f(x) = -(hx^3 - ix^2 + jx - 1) \qquad (11)$$

This latter selection environment is represented as the dashed line in Figure 3b, for the case where all parameters are equal to one. As long as the selections operate synchronously, the global and the stable points of inflection coincide (in a so-called "saddle point"). The historically stabilized system can then be identical with the global one. One can perhaps consider such a system as containing an identity since the global optimum coincides with the localized one.

Figure 3b shows also the configuration when stabilization and globalization no longer operate with the same parameter values. In other words, the selections have become asynchronous, as in systems containing functional differentiation in subroutines. In this case one expects the curve to show both a maximum and a minimum. At the minimum the system is stabilized, but at the maximum it can be considered as meta-stable. A bifurcation is thus induced because the system can go backward (to the stabilization of a trajectory) or forward (to globalization into a next-order regime).

While the system develops historically along a trajectory, the flux tends to move the system towards the other basin of attraction. This attraction is caused by the possibility to communicate in one more dimension, and thus to process more complexity in the newly emerging configuration (Turing, 1942). The development of this additional degree of freedom is endogenous to an evolving system (Dubois, 1998b).

The interactions between the various subdynamics make the system complex and result in the expectation of continuous transitions between provisional (that is, local) stabilizations and globalization at the systems level. Stabilization can be considered as a result of integration (e.g., by organization), while differentiation among the self-organizing fluxes can be expected to prevail when diffusion is no longer controllable at a center. However, the distinction between these two subdynamics—integration and differentiation—remains analytical; in the social system, the organization of communications at interfaces and the self-organization of the fluxes of communication can be expected to concur, since the global system cannot be historically manifest without some form of stabilization as a retention mechanism developing at the same time (Luhmann, 2000).

**6. Hyper-incursion**

How would these subdynamics of stabilization and globalization operate in the case of incursive selection mechanisms? The generation of meaningful information by the first incursion at each moment (Selection 1 in Figure 1) remains a necessary condition for the further development of the system. However, the one selection does not have to be attended by the other two continuously. Each two selection mechanisms can be expected to interact. A third incursive mechanism can additionally interact with the interaction term between two incursive selections terms, and this can generate a next-order incursion or hyper-incursivity.

A hyper-incursive equation no longer refers necessarily to a historical realization because it is based on recursivity in the anticipations and interactions among them. The most radical among the hyper-incursive equations reflects this orientation to the future as follows:

$$x_t = a x_{t+1} (1 - x_{t+1}) \qquad (12)$$

This system no longer contains any reference to its previous state $x_{t-1}$ or its current state at $t = t$ (that is, $x_t$), but the emerging state is considered as a function of the future state. When both expectations are interfaced on the basis of incursions of differentiated subsystems, we can use Equation 8 to formulate:

$$x_{i,t} = a x_{i,t+1}(1 - x_{j,t+1}) \qquad (13)$$

Elsewhere, it could be shown that only a historical variation can be produced when incursions are interfaced *recursively*. However, when the two different incursions are interfaced *incursively*, hyper-incursion can also be expected (Leydesdorff, 2006c).

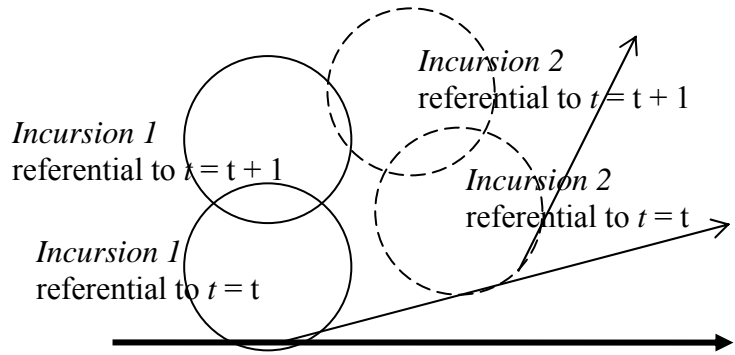

**Figure 4**: Hyper-incursivity at the interface emerging at $t = t + 1$ between two incursive routines.

While each incursion generates meaningful information and thus becomes both organized historically (with reference to $t = t$) and reflexive (with reference to $t = (t + 1)$), this hyper-incursive equation cannot be organized at $t = t$ because it contains only references to $t = (t + 1)$. The hyper-incursion builds on the interaction between references to future states in two underlying incursions (Figure 4).

Since this hyper-incursion no longer contains a reference to the current (historical) time, an additional subdynamic is needed in order to make its results relevant in historical time. Luhmann (2000) hypothesized that self-organization among the differently coded fluxes of communication—which generate only expectations—can be brought under organizational control by making decisions. Let us first see how this works in terms of the equations.

Equation 12 can be rewritten as follows:

$$x_t = ax_{t+1}(1 - x_{t+1}) \tag{12}$$

$$x_t = ax_{t+1} - ax_{t+1}^2$$

$$ax_{t+1}^2 - ax_{t+1} + x_t = 0$$

$$x_{t+1}^2 - x_{t+1} + x_t/a = 0$$

For $a = 4$, $x_{t+1}$ is defined as a function of $x_t$ as follows:

$$x_{t+1} = \tfrac{1}{2} \pm \tfrac{1}{2} \sqrt{(1 - x_t)} \tag{14}$$

Depending on the plus or the minus sign in the equation, two future states are generated at each time step. Since this formula is iterative, the number of future states doubles with each next time step. After $N$ time steps, $2^N$ future states would be possible. (For $N = 10$, the number of options is more than one thousand.)

Dubois (2003, at p. 115) specified a decision function $u(t)$ that can be added for making a choice between two options:

$$u(t) = 2\,d(t) - 1 \tag{15}$$

The author focused on $u = +1$ for the decision $d = 1$ (true) and $u = -1$ for the decision $d = 0$ (false). In a social system, however, more choices than these two extremes are possible. Social systems operate in a distributed mode with a probability distribution of preferences. This distribution contains an uncertainty. In distributed systems, decisions can be organized and codified into decision rules (Bertsekas & Tsitsiklis, 1989). Luhmann (2000) elaborated how organizations can be considered as the results of the codification of making decisions. The stabilization of decision rules generates the institutional layer of the social system in which decision-makers can develop routines.

Note that the decisions or decision rules[6] do not determine the hyper-incursive dynamics of the system at the regime level, but only guide these dynamics historically along trajectories (Dosi, 1982). However, the hyper-incursion continues to create possible futures as interactions among expectations generated on the basis of the underlying incursions. Decisions don't have to be taken at each step. The distribution of decisions and non-decisions changes the historical conditions by organizing the system, that is, by closing the circle in Figure 1. Without this historicity, the interfacing of expectations would remain in a virtual realm.

The horizon of possible meanings itself cannot be further developed historically without social realizations. The relevant decisions, however, are no longer taken about historical facts, but about expectations. The social system can become increasingly knowledge-based by this orientation towards the future because anticipations are continuously fed into its historical development. The expectations are based on the historical situation, but proliferating at a higher speed. In this configuration, the historical realizations may result from the interfacing of expectations as a base other than—but in interaction with—the historical organization of "reality" (Equation 3). The cycling among the selections adds meaningful information to the information flux which develops with the time axis. By feeding back on the time-axis, the meaning-processing system increasingly reconstructs the history of the system (cf. Hellsten *et al.*, 2006).

In other words, decisions specify instantiations among the available options when sets of expectations are interfaced. The organization of decisions along a historical axis potentially stabilizes a trajectory within a phase space of possible expectations. Social order is made contingent by coupling expectations about other realities with the historical reality of observable actions and institutions. The latter are increasingly reconstructed as the results of interactions among expectations.

---

[6] A codified decision rule could function as another incursive mechanism (Leydesdorff, 2006b, at pp. 147 ff.).

Historically, one may be inclined to consider the actions and institutions as the causes of the expectations because the former precede the latter. However, anticipations are based on the perspective of hindsight. When the feedback is codified and the codification reinforced by global horizons of meaning, the reconstructive arrow (from the perspective of hindsight) can after some time become more important than the forward arrow of historical development. While the historical construction continues to feed into the cycle bottom-up, control in communication systems thus tends to become increasingly top-down when the cycle (of Figure 1) can virtually be closed and systematically feed back as a next-order subdynamics. Unlike "natural" preferences grounded in the historical direction of the time axis, informed decisions guide development into an increasingly knowledge-based order given the contingencies of historical situations as relevant contexts.

**7. Mutual anticipation in social systems**

Two other hyper-incursive models follow from the logistic equation:

$$x_t = ax_t(1 - x_{t+1}) \qquad (16)$$

$$x_t = ax_{t+1}(1 - x_t) \qquad (17)$$

- Equation 16 evolves into $x = (a - 1)/a =$ Constant for each value of $a$. I submit as an appreciation that this evolution towards a constant through anticipation can be considered as modeling the self-reference or the expectation of an identity. Identity is based on the expectation of continuity of the "self" in the next stage. Like individuals, an organization can be expected to develop an identity;
- Equation 17 evolves into $x_{t+1} = (1/a) [x_t / (1 - x_t)]$. This routine formalizes the reflexive operation: when $x_t > [a / (1 + a)]$ a pulse is generated which first overshoots the value of one (in a virtual domain of possible expectations), but then generates a negative value (Figure 5). The negative value provides a mirror image of the representation at a specific moment in time, and thus allows for the reflection.

Reflection enables us to bounce a communication between the subdynamics of communication systems, and additionally between communication and consciousness because of the "interpenetration" between these two systems (Luhmann, 1988, 1991).

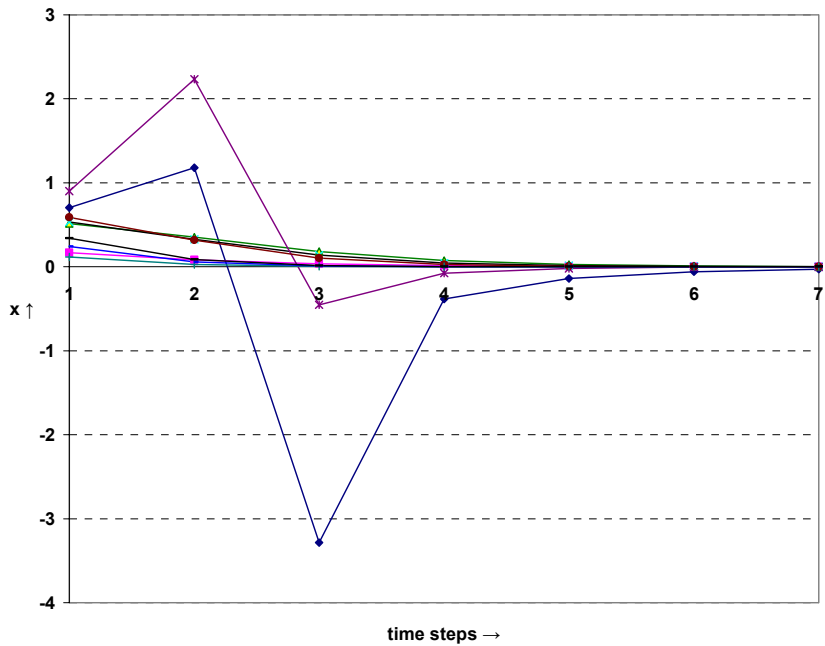

**Figure 5:** Simulation of Equation 18: the value of $x$ at $t = 1$ is drawn randomly ($a = 4$).[7]

In other words, the three hyper-incursive equations specify the fundamental processes of meaning-processing: identity formation (Eq. 16), reflexivity (Eq. 17), and the mutual expectation of expectations in the double contingency between reflexive (sub)systems (Eq. 12). While identity and reflexivity are more commonly defined in social theory, the expectation of expectations among systems may require additional explanation. Let me develop this in analogy to the concept of "double contingency" in the relation between *Ego* and *Alter*, but at the level of differently coded subsystems of society.

It was discussed above how "double contingency" has been theorized in sociology as an interaction between *Ego* and *Alter*. *Ego* and *Alter* are used in order to emphasize that in the relation between two human beings the *Ego* knows the *Alter* to be a reflexive *Cogito* who entertains expectations. For the *Ego* this means, that one's own reflexivity is reinforced in the encounter. The *Ego* in the present ($x_t$) no longer has a reference to itself other than in a future state ($x_{t+1}$), that is, as an *Alter Ego*.

---

[7] For $a = 4$, the pulse is generated for values of $x_t > 0.8$.

The orientation of providing meaning with reference to other possible meanings thus constitutes the social world as an order different from psychological orders (Husserl, 1929; Schutz, 1932, 1952, and 1975). Habermas (1987) criticized Luhmann's social systems theory for having replaced the traditional subject of transcendental philosophy (that is, the reflexive *Ego*) with the concept of a social system containing reflexivity (Leydesdorff, 2000). Let me turn Habermas's objection into a heuristics for understanding meaning-processing in a social system. What might be considered as the analogue of subjects at the level of the social system?

Subjects can entertain relations and experience double contingency in these relations, but let us follow Luhmann (1984, 1994, 1997) by assuming that there is only a single social system of communications—that is, society—containing subsystems. One can expect differently codified *sub*systems of communication to be reproduced under the condition of functional differentiation. These different subsystems are interfaced in historical organizations, but additionally they develop expectations which contain references to other possible expectations. Given this feedback, the reference of an organization to the time horizon may evolve from developing with history into developing from expectations about the future, and thus to becoming increasingly knowledge-based. Under the condition of functional differentiation, the (second) contingency along the time axis can be expected to change historically in the direction of an increasing orientation towards the future.

Unlike the self-organizing fluxes of communication in social systems, historically organized systems contain additionally a reference to themselves as an identity (Equation 16). (Individuals can be considered as the smallest units which organize the fluxes.) The social order of expectations remains structurally coupled to historical formats of organization because the non-linear dynamics of meaning-processing cannot be instantiated otherwise. The self-organization of fluxes of meaning-processing in different directions is not itself historically observable, but it can be made perceptible after a reflexive turn (Husserl, 1929; Luhmann, 1984). This coupling between the social system of communications and the agents carrying it is provided by the reflexivity formalized in Equation 17. Because of this reflexivity the social system can be entertained by individuals as a notional order, that is, an order with the epistemological status of a hypothesis.

*Vice versa*, the individuals and institutional agents are reflected in the social system as addresses or nodes. Since both systems can be expected to contain the reflexive operation expressed in Equation 17, a coevolution ("mutual shaping") between the social and individuals can also take place. However, this coevolution can be interrupted when expectations are further codified (e.g., stabilized) on either side beyond the control of the other side. On the side of social systems, an eigendynamics of codifications may lead to alienation. On the side of individuals, the recursive application of the reflexive operation provides discretionary room for private thinking and tacit knowledge.

In summary, functional differentiation changes the social system so that a double contingency between functionally differentiated, that is, differently coded, subsystems of communication becomes possible. Each of the subsystems can be expected to generate its own set of expectations. As these sets of expectations are made relevant to one another, their identity is no longer given, but has to be organized. Organization means in this case limiting the range of options by making decisions about the relevant expectations. For example, in the case of innovation one combines specific market perspectives with technological options. From this perspective, the individual (e.g., the entrepreneur) can be considered as the minimal carrier of decisions because decision-making itself can also be a social process.[8] Series of decisions shape a contingent trajectory which guides the self-organizing dynamics in a second contingency among expectations.

## 8. Conclusions

---

[8] Schumpeter (1934) stressed the role of the individual entrepreneur in innovations ("Schumpeter Mark I"). Schumpeter (1943) developed "Mark II" in which innovation was envisaged as a more routinized process within large firms.

The functional differentiation of society in terms of different codes of communication provides the social system with an additional mechanism of anticipation because of the asynchronicity in the updates among the different subsystems. In organized systems and hierarchies, the transversal updates at subsystemic levels are synchronized at the top with reference to an external environment, but a functionally differentiated system has replaced this option for survival by considering the subsystems as selection environments for one another.

The transversal generation of meaning among the subsystems can be recombined with the longitudinal generation of meaning and thus provide a field of possible meanings. The resulting selection environments are no longer linear, but can be shaped in terms of trajectories and regimes. Two selection environments can be expected to co-evolve along a trajectory in a process of mutual shaping; three selection environments provide sufficient complexity for generating both the possibility of local stabilizations (niches) with their respective trajectories, and meta-stabilization or globalization at the regime level. There is no prescribed order: the different meaning-providing mechanisms both interact in exchanges of meaning and operate recursively with reference to previous constructions of meaning. Thus, a non-linear dynamics of meaning processing in social systems can be generated and reproduced.

The second-order incursion upon the anticipatory terms of the first-order incursions provided us with a hyper-incursive equation which modeled a *third* selection mechanism. The hyper-incursive mechanism operates upon the basis of the anticipatory models provided by the incursive routines and thus abstracts from the historically prior constructions which were modeled by the incursive models. However, the uncertainty which is generated by this hyper-incursive routine forces the system to be (re)organized periodically like an update mechanism. I suggested that the minimum unit for this historicity is the individual.

At the level of individual interactions, hyper-incursivity is also available as a "double contingency": the meeting of the other who entertains a similar, but potentially different set of expectations. "Double contingency" can thus be considered as a transversal encounter between different sets of meanings, but in the case of individuals whole systems (agents) meet and update, and not subsystems. Subsystems of communication operate at a different level and can only reproduce a double contingency towards one another if the functional differentiation among them is warranted in terms of codes developing along sufficiently different axes.

The social system shares with psychological systems the other two hyper-incursive operations (identity formation and reflexivity). Unlike individuals, the social system is able to generate within the system a non-linear dynamic among *three* selection mechanisms: (1) selections from a variety of global horizons of meaning, (2) historical realizations in organizations at each moment, and (3) the transformation of the latter by the need to recombine expectations based on previous realizations over time. The three selections operate upon one another: when the configuration allows for interaction of a third selection mechanism with the interaction between two incursive selections, hyper-incursion becomes a possibility, and in this case the systems can be expected to avalanche in the strongly anticipatory regime of a knowledge-based society. In a knowledge-based order, decisions are increasingly informed by expectations.

Let me finally note that the argument in this paper would indicate that the social system as a strongly anticipatory one would *not* be able to construct its own next stage in the longer run without historical decision-making by agency. Thus, social systems can be considered as *semi*-autonomous (Collier, 2005). In addition to structural coupling as systems, interpenetration between consciousness and communication provide these meaning-processing systems with an operational coupling (Luhmann, 1988, 1991).

While Dubois (2002) showed that a strongly anticipatory system is able to predict the development of a complex system given a fixed relation between the modeling and the modeled system—e.g., in the case of master-slave models—the hierarchical relations can be expected to remain changing in social systems which are functionally differentiated because of asymmetries in the selections of subsystems upon each other. Decision-making structures (e.g., individuals) are a necessary condition for the historical development of the non-linear dynamics of meaning in social systems. The sequential decisions construct trajectories which carry the next-order cycle. The next-order cycle feeds back on the possible trajectories in its domain as a regime (Dosi, 1982; Maturana, 2000).

**Acknowledgement**

I thank John Collier and an anonymous referee for comments on an earlier draft.